\documentstyle[epsf,rotate]{article}

\newcommand{\MeV}{\mbox{MeV}}
\newcommand{\GeV}{\mbox{GeV}}

\begin{document}

\begin{flushright}
Liverpool Preprint: LTH \\
 \end{flushright}


\vspace{20mm}
\begin{center}
{\LARGE \bf An estimate of the chiral condensate\\ from unquenched 
lattice QCD.}\\[15mm]

 {\bf C. McNeile  \\
Theoretical Physics Division, Dept. of Mathematical Sciences, 
          University of Liverpool, Liverpool L69 3BX, UK 
 }\\[2mm]

\end{center}

\begin{abstract}
Using 
the parameters in the chiral Lagrangian obtained by MILC from their
unquenched lattice QCD calculations with 2+1 flavours of sea quarks, I estimate
 the chiral condensate. I obtain the result
$\langle \overline{\psi}\psi\rangle (2\; \GeV) / n_f$ =
$-(259 \pm  27  \; \MeV)^3$ in the $\overline{MS}$ scheme. 
I compare this value to other
determinations.
 \end{abstract}
%


\section{Introduction}

The spontaneous breaking of chiral symmetry 
plays an important role in the dynamics of 
low energy QCD. The non-zero value for the chiral condensate 
is caused by spontaneous chiral symmetry breaking.
The chiral condensate
is a basic parameter in the QCD sum rule approach to 
computing 
hadronic
quantities~\cite{Dosch:1997wb,Narison:2002hk} so a numerical
value from lattice QCD is a valuable check on that formalism.

There have been many quenched lattice QCD calculations that have
reported a value for the chiral
condensate~\cite{Becirevic:2004qv,Giusti:1998wy,Gimenez:2005nt,DeGrand:2001ie}.
The MILC collaboration~\cite{Bernard:2001av,Aubin:2004wf,Aubin:2004fs}
have been performing unquenched lattice QCD calculations with the most
realistic set of parameters used to date.  The results from MILC's
lattice calculation have been successfully compared against experiment
for many quantities that are stable to 
strong decay~\cite{Davies:2003ik}.
MILC's lattice calculations
use the improved staggered fermion action.
The method of performing unquenched calculations
with improved staggered quarks has potential problems
with non-locality (see~\cite{DeGrand:2005is} for
a review), however this problem has not
shown up in the comparison of results
currently computed against experiment.
The unquenched calculations use 2 light quarks in the sea
and one sea quark fixed at approximately the strange quark mass.
The data set from MILC has two lattice spacings (0.09 fm and 0.125 fm).
All the volumes were larger than 2.5 fm. The lightest 
pion mass used in MILC's calculation is 250 \MeV.

Although the data from MILC's unquenched calculations has been
used do much important phenomenology, it has not been used
to estimate a physical value for the 
chiral condensate. In this paper, I estimate the 
chiral condensate from the latest published MILC data. 

\section{Extracting the chiral condensate using chiral perturbation theory.}

In an extensive calculation~\cite{Aubin:2004fs}
the MILC collaboration 
fitted the squared masses of pseudo-scalar mesons 
and pseudo-scalar decay constants
to 
the expressions from chiral perturbation 
theory~\cite{Lee:1999zx,Bernard:2001yj,Aubin:2003uc,Aubin:2003mg}
\begin{equation}
M_{PS}^2 / ( m_x + m_y) = \mu ( 1 + ... ) 
\label{eq:PSmasquark}
\end{equation}
where $M_{PS}$ is the mass of the pseudo-scalar
meson made of quarks $x$ and $y$ with masses $m_x$
and $m_y$ respectively. In equation~\ref{eq:PSmasquark}
the dots represent higher order terms that MILC included 
in the fits.

Chiral perturbation theory relates the chiral condensate 
$\langle \overline{\psi} \psi \rangle / n_f$
to $\mu$
\begin{equation}
\langle \overline{\psi} \psi \rangle / n_f = - \frac{1}{2} \mu f^2
\label{eq:master}
\end{equation}
where $f$ is the pion decay constant in the chiral limit.
I use the normalisation of the axial current such that
the physical pion decay constant is 131 \MeV.
When there are no correction terms in equation~\ref{eq:PSmasquark},
these two equations are known as the Gell-Mann-Oakes-Renner 
formulae (GMOR). Strictly speaking the GMOR relation 
can only be used to extract the chiral condensate if 
there are no higher order corrections to 
equation~\ref{eq:PSmasquark}.
Chiral perturbation theory is the 
generalisation of GMOR to higher orders in the 
pion mass. The extraction of the chiral condensate from 
the parameters of the chiral Lagrangian obtained by
fits to lattice data in the continuum large volume
limits may be an empirical approach, but 
I believe it is valuable.

In equation~\ref{eq:master},  $\langle \overline{\psi} \psi \rangle$
is the sum of the chiral condensates for each sea quark.
In the appropriate limit where the masses of all
the sea quarks go to zero in the infinite volume limit,
then the chiral condensate of each sea 
quark is the same~\cite{Scherer:2002tk,Vafa:1983tf}.

Another technique to extract the chiral condensate from
lattice QCD calculations is to compute the scalar 
correlator 
directly~\cite{Becirevic:2004qv,Gimenez:2005nt,DeGrand:2001ie}. 
The results are then extrapolated 
to the zero quark mass limit.
In a quenched lattice QCD calculation,
Be\'cirevi\'c\ and Lubicz used both the GMOR and 
scalar correlators to extract
a consistent result for the 
chiral condensate~\cite{Becirevic:2004qv}.

The use of the parameters from the chiral 
Lagrangian that have been fitted from
lattice data to estimate the chiral condensate 
may miss some interesting physics. 
Stern and 
collaborators~\cite{Descotes-Genon:2003cg,Descotes-Genon:1999uh,Descotes-Genon:2004iu}
have proposed a scenario where the chiral condensate
for three flavours is very small, so the higher order
terms in equation~\ref{eq:PSmasquark} contribute the 
majority of the meson mass. The possibility that the 
two flavour chiral condensate is small has been 
ruled out 
in the theory with 
$n_f=2$
by comparison of chiral perturbation theory
with $\pi\pi$ scattering~\cite{Leutwyler:2001hn}.

\section{Extracting the result} \label{eq:numericalAnaly}

MILC have recently reported a fit of the numerical 
data for light pseudo-scalar meson masses and 
decay constants to expressions derived from 
staggered
chiral
perturbation theory~\cite{Aubin:2004fs}.
A simultaneous fit was done to the data 
at the two different lattice spacings.
The fit functions for the masses and decay constants
used NNLO analytic terms, as well as lattice
artifact terms arising from working at fixed lattice spacing
and fixed lattice volume.

I use equation~\ref{eq:master} with the results
from MILC's extensive fits to chiral perturbation
theory. One important issue is the renormalisation of
equation~\ref{eq:master}. The MILC collaboration used 
a conserved axial current, so no renormalisation factor is needed
for the pion decay constant in the chiral limit ($f$). The 
$\mu$ term does need to be renormalised. The $\mu$ term
is renormalised with $Z_S$ that is related to the renormalisation
of the mass via $Z_S = \frac{1}{Z_m}$. The $Z_m$ renormalisation
factor computed by the HPQCD, MILC, and UKQCD 
collaborations
 is reported in~\cite{Aubin:2004ck}.
\begin{equation}
Z_m(\Lambda)  = 
\frac{1}{u_0}
\left( 1 + \alpha \left( b - \frac{4}{3 \pi} - \frac{2} {\pi} \ln( a \Lambda) 
\right) \right)
\end{equation}
where $\alpha$ is the QCD coupling, $b = 0.5432$, and
$u_0$ is the tadpole factor. I always quote numbers at
the scale 2 GeV in the $\overline{MS}$ scheme.

The MILC analysis~\cite{Aubin:2004fs} 
is a combined fit to data at two lattice spacings.
The convention for the quark mass renormalisation is
to convert the quark masses to the lattice scheme 
on the fine lattice. Hence, the $Z_S$ factor for the 
quark masses on the fine lattice must be used to convert
$\mu$ into the $\overline{MS}$ scheme at a scale of 2 \GeV.
Using the numbers quoted by MILC~\cite{Aubin:2004fs} 
I get $Z_m(2 \; \GeV)$ = $1.195$.
The two loop computation of $Z_m$ in lattice perturbation
theory is underway~\cite{Trottier:2003bw}. 
A non-perturbative estimate of $Z_m$, using 
similar techniques to those used by JLQCD~\cite{Aoki:1999mr}
to renormalise the quark mass with Kogut-Susskind
fermions, would be useful.
I use the 
estimate of 9\% from MILC~\cite{Aubin:2004fs} 
for the error due to the 
truncation of the perturbative series.

One advantage of the MILC collaboration's calculation
is that a consistent lattice spacing is obtained
from many different quantities that are stable against 
strong decay~\cite{Davies:2003ik}. 
The chiral condensate involves the 
third power of the lattice spacing, so any errors in the 
choice of scale are amplified.
I used MILC's value~\cite{Aubin:2004wf}
$r_1 = 0.317(7)$ fm to convert the $\mu$ and $f$
parameters into physical units.

MILC's analysis~\cite{Aubin:2004ck} of their  data
used a number of fits that included various subsets
of their data. In table~\ref{tab:MILCcuts}, I compute
$\langle \overline{\psi}\psi\rangle(2 \;\GeV) / n_f$ 
using equation~\ref{eq:master} and the perturbative
matching factor with the 
values for $\mu$ and $f$ in table IV
of~\cite{Aubin:2004fs}. The results for the three main fits
(called A, B and C) are in table~\ref{tab:MILCcuts}.
The errors for the chiral condensate are dominated 
by the error on the pion decay constant in the chiral limit 
($f$).
\begin{table}[tb]
  \centering
  \begin{tabular}{|c|c|c|c|c|} \hline
Fit  & Points  &  Coarse & Fine  & $\langle \overline{\psi}\psi\rangle(2 \;\GeV) / n_f$   \\ \hline
A & 94 & $m_x + m_y < 0.40m_s'$ & $m_x + m_y < 0.54 m_s'$ & $(269 \pm
17  \; \MeV)^3$
\\ \hline
B & 240 & $m_x + m_y < 0.70m_s'$ & $m_x + m_y < 0.80m_s$ & $(250  \pm 21 \; \MeV)^3$
\\ \hline
C & 316 & $m_x + m_y < 1.10m_s'$ & $m_x + m_y < 1.14m_s'$ & $(249 \pm 15 \; \MeV)^3$ \\ \hline
  \end{tabular}
  \caption{Results for chiral condensate
in the $\overline{MS}$ scheme at a scale of 2 \GeV\ for
various fits that MILC did to their data. The coarse and fine 
columns correspond to the mass ranges used in the fits with
the data on the coarse and fine lattice spacing.
The $m_s'$ is
the mass of the strange sea quark in MILC's unquenched
calculation. The points column is the number of data used in the fit.}
\label{tab:MILCcuts}
\end{table}

MILC use combinations of the results from the fits A,B, and C to
estimate the central values and the systematic errors.
I take the average of the result for fit
A and B as the central value.
\begin{eqnarray}
\langle \overline{\psi }  \psi \rangle (2 \; \GeV) / n_f & = & -0.018(5) \; \GeV^3
\\
& = &  -(259 \pm 27 \; \MeV)^3
\end{eqnarray}

I now discuss the important issue as to whether 2+1=3.
There are two main possibilities, that the chiral perturbation
theory analysis is sensitive to the three flavour
chiral condensate or to the two flavour chiral condensate with
one sea quark fixed at the strange quark mass. In the 
latter case the three flavour chiral condensate could be 
extracted with the help of chiral perturbation 
theory~\cite{Gasser:1984gg}.
In the MILC calculation the mass of the strange quark is 
fixed. However, the data is analysed with three
flavour chiral perturbation theory. The mass of the
strange quark was slightly incorrect on the coarse
lattice, so some small extrapolation in the data is
done for the strange quark mass. 
It is hard to give a definitive answer to the
flavour dependence of the condensate extracted in this paper
without further analysis.

I could not find any recent calculations of the 
chiral condensate from unquenched lattice QCD calculations
with two flavours of sea quarks. Early work is 
reviewed in~\cite{Gupta:1996sa}. I have used the chiral
perturbation theory approach to extract the the chiral
condensate from a recent unquenched lattice QCD calculation 
by the JLQCD collaboration~\cite{Aoki:2002uc}.
This lattice calculation was done at a fixed
lattice spacing of 0.089 fm. The lightest sea quark mass
was half the strange quark mass.
The axial vector current used in JLQCD's calculation needs
to be renormalised. I used tadpole improved perturbation
theory
with a simple boosted coupling to estimate the 
required renormalisation (using the summary of results in the 
appendix of~\cite{Bhattacharya:2000pn}). The numerical value of 
renormalisation factor is 0.45, so some kind of 
non-perturbative renormalisation is required for 
a definitive answer, hence the error for JLQCD estimate is 
unreliable. D\"urr~\cite{Durr:2002zx} has previously noted
the problems with extracting the chiral condensate from 
unquenched calculations done by the CP-PACS and UKQCD 
collaborations.

\section{Conclusion and comparison to other work} \label{eq:conc}

In table~\ref{tab:CSummary}, I compare my
analysis of the MILC and JLQCD data to a selection of recent lattice
results for the chiral condensate.

\begin{table}[tb]
  \centering
  \begin{tabular}{|c|c|c|c|} \hline
Group  & $n_f$ & $\langle \overline{\psi}\psi\rangle(2 \;\GeV) / n_f$  & 
$\langle \overline{\psi}\psi\rangle(2 \;\GeV) / n_f$   \\ \hline
This work, MILC &  2+1 & $ -0.017(5) \GeV^3$  & $-(259 \pm 27  \; \MeV)^3 $  \\
This work, JLQCD &  2  & $ -0.009(1) \GeV^3 $  & $-(209 \pm 8  \; \MeV)^3 $  \\
Be\'cirevi\'c\ \& Lubicz~\cite{Becirevic:2004qv} & 0 &   &
$-(273 \pm 19 \; \MeV)^3$  \\ 
Be\'cirevi\'c\ \& Lubicz~\cite{Becirevic:2004qv} & 0 &   &
$-(312 \pm 24 \; \MeV)^3$  \\ 
Giusti et al.~\cite{Giusti:1998wy} & 0 & $-0.0147(8)(16)(12) \; \GeV^3$ &  
$-(245(4)(9)(7) \; \MeV )^3$
\\
Gimenez et al.~\cite{Gimenez:2005nt} & 0 &  & $-(265 \pm 5 \pm
22 \;\MeV)^3$   \\
Hernandez et al.~\cite{Hernandez:2001yn} & 0 &  &  $-(268(12) \; \MeV )^3$ \\
DeGrand~\cite{DeGrand:2001ie} & 0 &  &  $-(282(6) \; \MeV )^3$ \\
Giusti et al.~\cite{Giusti:2001pk} & 0 &  &  $-(267(5)(15) \; \MeV )^3$ \\
Blum et al.~\cite{Blum:2000kn} & 0 &  &  $-(256(8) \; \MeV )^3$ \\
\hline
  \end{tabular}
  \caption{Results for chiral condensate
in the $\overline{MS}$ scheme at a scale of 2 \GeV. }
\label{tab:CSummary}
\end{table}

Giusti et al.~\cite{Giusti:1998wy} note that their
numbers are comparable to estimates of the chiral condensate from
sum rules~\cite{Dosch:1997wb,Narison:2002hk}. 
The first entry in table~\ref{tab:CSummary} from 
Be\'cirevi\'c\ \& Lubicz comes from a GMOR analysis and the 
second is from the pseudo-scalar vertex.
Pennington~\cite{Pennington:2002nb}
reviews various calculations of the chiral condensate
from lattice, and sum rules
and estimates the size of the 
the chiral
condensate to be $\sim -(270 \; \MeV)^3$.
Jamin~\cite{Jamin:2002ev} obtained a value for 
the chiral condensate of 
$\sim -(267 \pm 16 \;\MeV)^3$
from QCD sum rules.

From table~\ref{tab:CSummary}, I note that the 
result from MILC is essentially consistent with the other
results. 
Descotes et al. have argued that the chiral condensate with
three sea quarks should be less than that from 
QCD with two light sea quarks~\cite{Descotes-Genon:1999uh}.
Given the assumptions in this analysis it does not look as
though the  chiral condensate has a strong dependence
on the number of quarks in the sea.
The Columbia~\cite{Brown:1992fz,Mawhinney:1996jk} group 
claimed to see a reduction in
chiral symmetry breaking from unquenched calculations with
0, 2, and 4 flavours of sea quarks, but the analysis 
of their data was complicated by finite size effects.
From lattice QCD calculations Iwasaki et al.~\cite{Iwasaki:2003de}
find that the theory becomes deconfined for $n_f > 6$.

From equation~\ref{eq:PSmasquark}, the higher value for the chiral 
condensate is correlated with smaller quark masses.
This trend was noted by Gupta and Bhattacharya in a review
of lattice data before 1997~\cite{Gupta:1996sa}.
Although when higher order mass corrections  are included
in equation~\ref{eq:PSmasquark}
 this is not
so obvious. 

From the same data set, the MILC, HPQCD and UKQCD
collaborations~\cite{Aubin:2004ck} have obtained the mass of the strange quark
to be $m_s^{\overline{MS}}$(2 \GeV) = 76(0)(3)(7)(0) \MeV,
This value for the strange quark mass is low relative
to other determinations~\cite{Lubicz:2000ch,Wittig:2002ux,Rakow:2004vj},
however, this calculation is the first large scale lattice
calculation with 2+1 flavours of dynamical light quarks.
The only other group to have published results 
for the strange quark mass from unquenched 
lattice QCD calculations with $2+1$ flavours of 
sea quarks is the JLQCD/CP-PACS collaboration. 
At a fixed lattice spacing, they obtain $m_s(2 \GeV)$ 
between 80 and 90 \MeV~\cite{Ishikawa:2004xq}. The 
JLQCD/CP-PACS collaboration are planing to compute the 
strange quark mass at other lattice spacings to do
a continuum extrapolation.

Unfortunately, the data in table~\ref{tab:CSummary}
are not precise enough to understand the systematics
of quark mass determinations  between lattice QCD calculations
with 2 and 3 flavours of sea quarks. A reduction
in the size of errors in the estimates of the chiral
condensate from lattice calculations with a different
number of sea quarks would help 
compare the results for quark masses from different
calculations.

The main theoretical concern with unquenched 
calculations with improved staggered fermions is that 
the formalism requires taking the fourth of 
the determinant that controls the sea quark dynamics. 
There have been a number of theoretical
papers on this 
topic~\cite{Durr:2003xs,Durr:2004as,Durr:2004ta,Follana:2004sz,Shamir:2004zc,Adams:2004mf,Maresca:2004me}
(the issues are adroitly explained by DeGrand~\cite{DeGrand:2005is}).
None of the theory papers on the locality of improved staggered fermions
have satisfactorily resolved the issue for QCD.
One of the main tests 
of the fourth root trick is comparison of the lattice data
with the results from
chiral perturbation theory~\cite{Aubin:2004fs}, 
hence it is important to fully
understand all aspects of chiral perturbation theory applied
to the MILC data. 
Crosschecks on the chiral perturbation theory analysis of MILC's data
are also very valuable, because of the important phenomenology
extracted
from their work.
An independent computation of the magnitude of the chiral
condensate using different correlators would be a useful
crosscheck~\cite{Hands:1990wc,Salina:1991wr} on the estimate from the chiral
perturbation theory fits. MILC do study the chiral condensate 
in their work on finite temperature~\cite{Bernard:2004je}.

\section{Acknowledgements}
I thank Chris Michael for discussions and for reading the paper.
Thanks to Claude Bernard for an explanation of the quark mass conventions
in the MILC data.


\end{document}